\documentclass{svjour2}                    
\smartqed  
\usepackage{graphicx}
\usepackage{amsmath,amssymb}
\usepackage{psfig}
\newcommand \be{\begin{eqnarray}}
\newcommand \ee{\end{eqnarray}}
\newcommand {\p}[1]{\partial_{#1}}
\newcommand \ba{\begin{align}}
\newcommand \eea{\end{align}}
\newcommand \s{{\backslash\hspace{-1ex}i
}}

\begin{document}

\title{Asymmetric Bethe-Salpeter equation for pairing and condensation
}


\author{Klaus Morawetz         
}


\institute{K. Morawetz \at
              M\"unster University of Applied Science,
Stegerwaldstrasse 39, 48565 Steinfurt, Germany
           \and
\at
International Institute of Physics (IIP)
Federal University of Rio Grande do Norte
Av. Odilon Gomes de Lima 1722, 59078-400 Natal, Brazil
\and
\at
Max-Planck-Institute for the Physics of Complex Systems, 01187 Dresden, Germany\\
              Tel.: +49 2551 962 411\\
              Fax: +49 2551 962 811\\
              \email{morawetz@fh-muenster.de}           
}

\date{Received: date / Accepted: date}

\maketitle

\begin{abstract}
The Martin-Schwinger hierarchy of correlations are reexamined and the three-particle
correlations are investigated under various partial summations. Besides the
known approximations of screened, ladder and maximally crossed diagrams the
pair-pair correlations are considered. It is shown that the recently
proposed asymmetric Bethe-Salpeter equation to avoid unphysical repeated
collisions is derived as a result of the hierarchical dependencies of
correlations. Exceeding the parquet approximation we show that an asymmetry
appears in the selfconsistent propagators. This form is superior over the
symmetric selfconsistent one since it provides the Nambu-Gorkov equations and
gap equation for fermions and the Beliaev equations for bosons while from the
symmetric form no gap equation results. The selfenergy diagrams which account
for the subtraction of unphysical repeated collisions are derived from the
pair-pair correlation in the three-particle Green's function. It is suggested to distinguish between two types of selfconsistency, the channel-dressed propagators and the completely dressed propagators, with the help of which the asymmetric expansion completes the Ward identity and is $\Phi$-derivable. 

\keywords{Many-body theory \and Correlations in Nonequilibrium \and Bethe-Salpeter equation}
 \PACS{71.10.-w \and 
05.30.-d \and 
74.20.-z \and 
05.60.Gg \and 
67.85.-d \and 
82.20.-w  
}
\end{abstract}

\section{Introduction}
Though pairing and condensation phenomena like superconductivity belong to
the most exciting and long studied effects, their many-body theoretical
treatments lacks often a systematic
justifications. For example a particle-number non-conserving assumption is
often accepted by anomalous propagators in order to obtain the desired
structure of Green's functions. Systematic many-body approximations like the
ladder T-matrix \cite{KM61,BSI74,BSI75,MBL99,HCCL07} leading to the Bethe-Salpeter
equation had to face the puzzle that the selfconsistent propagators describing
the many-body medium have to be assumed in asymmetric form in order to obtain
the gap equation. This was first observed by Kadanoff and Martin \cite{KM61}
and this approximation has been used mainly in the theory of superconductivity
\cite{MBL99,HCCL07,CHCL08}. For
an overview about different T-matrix approximations see \cite{CSTL05}. 
Though any T-matrix based on the Bethe-Salpeter equation becomes unstable at the critical 
temperature \cite{BSI75} where the T-matrix diverges, the corresponding
selfenergy fails to describe the 
superconducting gap \cite{Wild60} if we do not assume an asymmetry in the
internal propagators.

Recently a way has been proposed to obtain such asymmetry by  demanding that
particles should not interact with the same state repeatedly \cite{L08}.  Such
non-physical repetition leads to an incorrect description of the coherent part of the scattering \cite{Watson53}, and an incorrect 
single-particle Green function $G$. These non-physical repeated 
collisions destroy the superconducting gap in the energy spectrum of $G$.
Using the Soven scheme \cite{Soven67} the repeated collisions can be avoided 
\cite{L08} and the gap in the superconducting state can be described as well
as the normal state. This subtraction scheme can be written in a form
equivalent to the anomalous propagator approach such that these propagators appear as consequence of the theory and need not to be assumed ad-hoc \cite{M10}. Though in this way a physical understanding  of a
deficiency in selfconsistent many-body expansion has already
been achieved, it is desirable to see how a systematic expansion of correlations can account for such corrections. 

The necessity to introduce asymmetric propagators can be seen from another
physical system of dense correlated plasmas. There the motion of ions are
lowered by the dynamics of the screening cloud since a 
deceleration force appears which is caused by the deformed screening cloud
surrounding the charge. This effect known as
Debye-Onsager relaxation effect has been devoted much interest \cite{k58,KE72,e76,er79,r88,MK92,ER98,Mo00} which was first derived within
the theory of electrolytes \cite{DH23,O27,f53,FEK71,KKE66}.  
It turned out that only an asymmetric assumption \cite{Mo00} about
the mutual screening can correctly describe the Onsager result \cite{er79,r88,ER98}. 

Due to the widespread usage of T-matrix approximations ranging from nuclear phy\-sics for both
equilibrium \cite{KM93,Bo99,Bo02} and non-equilibrium \cite{bm90,LSM97}
problems, to the theory of moderately dense gases \cite{B69} and 
liquid $^3$He \cite{GBS76}, as well as to electron-electron 
correlations in molecules and solids
\cite{NIO06,Toulouse70,GKSB99,MPSSBR95,YSS07} it is desirable to clarify the
correct structure of the Bethe-Salpeter equation and to specify the conditions
under which the corresponding approximate forms can be used.

Here we will present a derivation of modified Bethe-Salpeter equations from
first-principles which shows indeed that the hierarchical structure of
correlation functions leads to such asymmetric selfconsistency as necessary to
describe pairing. We will present the derivation in terms of the causal
n-particle Green's functions. The cummulant expansion yields a hierarchy of
correlation functions which leads already to
the asymmetric result assumed by Kadanoff and Martin in binary correlation approximation. 

We will then proceed and
investigate the three-particle correlation and will select the processes
between two pairs of particles which allow to consider the repeated
collisions. We will show that the systematic cummulant expansion leads to the
correct subtraction in the channels with condensates called singular
channels. The idea of derivation follows here the centennial overview of
many-body approximations by Heinz Puff \cite{P79} which is suited for
nonequilibrium Green's function expansions. Therefore all outlined formalism
holds in nonequilibrium as well as equilibrium. The expansion scheme of
three-particle correlations will lead to the appearance of two different kinds
of selfconsistent propagators in each considered channel. These will be the channel-dressed and the completely dressed propagators.

The  outline of the paper is as follows. In the next chapter a 
scheme of Green's functions with respect to the cummulant
expansion is shortly presented. This is performed with the help of the variation technique
including an external potential in order to ensure the Ward-identity, $\phi$-derivability and
conservation laws which will be shown in the appendix. Then we consider in the third chapter the binary
approximation neglecting any three-particle correlation. This will lead to
the Kadanoff- and Martin approximation providing the correct gap equation. We
discuss then the repeated collisions and how it can be avoided by a proper
subtraction scheme. Though leading to the same gap equation the underlying
approximation is beyond the Kadanoff-Martin form. To justify this subtraction
scheme we consider in chapter IV the three-particle correlations. We
demonstrate that the expansion of three-particle correlations in different
channels leads to the distinction  of two kinds of selfconsistent propagators,
the channel dressed and the completely dressed one. Then we focus on
the correlations between two pairs. This provides a special contribution from
the pairing or condensation channel to the
different channels contained in the three-particle correlations. It turned out
that this selfenergy contribution provides exactly the subtracted scheme
discussed in chapter III and which was proposed in the literature. In the
appendix we proof the Ward identity and the $\Phi$-derivability in order to
ensure conservation laws. We
summarize and emphasize that the hierarchical dependence of correlation care
for the avoidance of unphysical repeated collisions itself. As a consequence the
T-matrix schemes used so far should be revised as soon as singular channels
appear like in phenomena of condensation or pairing in such a way that an asymmetry in
the selfconsistency is proposed.

\section{Decoupling scheme of correlations}

The n-particle causal Green's function 
\be
G(1,2,...,n,1',2',....,n';U)&=&{1\over i^n}{\langle T S \Psi_1...\Psi_n\Psi^+_{1'}...\Psi^+_{n'}\rangle \over \langle T S\rangle}\nonumber\\
 S&=&{\rm e}^{\int d1 \Psi_1^+\Psi_1 U(1)}
\label{green}
\ee
of a system under the influence of an external field $U(1)$, where numbers $1,2,...$ sign cumulative indices like space, time,... coordinates, can be formally represented by a generating functional
\be
&&G(1,2,...,n,1',2',....,n';U) 
=\left .\p{\eta_{n'}}...\p{\eta_{1'}}\p{\lambda_1}...\p{\lambda_n}{\cal G}[\lambda,\eta]\right |_{\lambda=\eta=0}
\ee
with 
\ba
&{\cal G}[\lambda,\eta]=1
+\sum\limits_{1}^n{1\over (n!)^2}\int d 1...dnd1'...dn' \lambda_n...\lambda_1 G(1...n';U)\eta_{1'}...\eta_{n'}
\label{1}
\end{align}
where $\eta$, $\lambda$ are Bose/Fermi-commuting auxiliary fields.

The Martin-Schwinger hierarchy \cite{MS59} coupling the one-particle Green's function to the two-particle one, 
\be
&&\left [i\p{t_1}+{\nabla_{r_1}^2\over 2m}-U(1)\right ] G(11';U)=\delta(1-1')
\mp i \int d\bar 1 V(1,\bar 1) G(1\bar1 1'\bar 1^+;U),
\label{MS}
\nonumber \\&&
\ee
is expressed in terms of this generating functional by
\be
&&\left [i\p{t_1}+{\nabla_{r_1}^2\over 2m}-U(1)\right ] \p{\lambda_1 }{\cal G}=\eta_1 {\cal G}
\mp i \int d\bar 1 V(1,\bar 1) \p{\eta_{\bar 1^+}}\p{\lambda_{\bar 1}} \p{\lambda_1} {\cal G}.
\label{2}
\ee
The upper sign denotes fermions the lower bosons hereafter.
It is now useful to introduce the correlated n-particle Green's function as the
cummulant expansion \cite{Fulde} due to a new generating functional ${\cal G}_c$,
\be
{\cal G}[\lambda,\eta]={\exp}^{{\cal G}_c[\lambda,\eta]}
\label{3}
\ee
with
\ba
&{\cal G}_c[\lambda,\eta]=
\sum\limits_{1}^n{1\over (n!)^2}\int d 1...dnd1'...dn' \lambda_n...\lambda_1 G_c(1...n';U)\eta_{1'}...\eta_{n'}.
\label{4}
\end{align}
The comparison of (\ref{3}), (\ref{4}) and (\ref{1}) with respect to the orders of $\lambda,\eta$ reveals the cummulant expansion. The first order reads
\be
G(11';U)=G_c(11';U),
\ee
i.e. the single-particle Green's function equals its correlated part. The second
order shows just the separation of the Hartree-Fock term from the two-particle Green's function
\ba
&G(121'2';U)=G_c(121'2';U)\nonumber\\&+G(11';U)G(22';U)\mp G(12';U)G(21';U).
\end{align}
The third order is given by all possible exchanges of three one-particle
Green's functions together with all possible exchanges of the two-particle correlated Green's function and a one-particle Green's function. 

Introducing (\ref{3}) into (\ref{2}) and notating the inverse Hartree-Fock Green's function with
\ba
&G_{\rm HF}^{-1}(12;U)=-i V(12) G(12;U)\nonumber\\
&\!+\!\left [i\p{t_1}\!+\!{\nabla_{r_1}^2\over 2m}\!-\!U(1)\!\pm\! i\int d\bar 1 V(1\bar1) G(\bar 1 \bar 1^+;U)\right ]\delta(1\!-\!2)
\label{GHF}
\end{align}
one can invert the differential equation (\ref{2}) into an integral equation
\be
\p{\lambda_{1}} {\cal G}_c&=&{\cal I} (1)
\label{invert}
\ee
with
\be
{\cal I}(1)&=&\int d\bar 1  G_{\rm HF}(1 \bar 1;U)\eta_{\bar 1} \mp i\int d\bar 1 d \bar 2  G_{\rm HF}(1 \bar 1;U) V(\bar 1 \bar 2) 
\nonumber\\ && \times
\left \{
\p{\eta_{\bar 2^+}}\p{\lambda_{\bar 2}}\p{\lambda_{\bar 1}} {\cal G}_c
\right . \nonumber\\ && \left .
+
[\p{\eta_{\bar 2^+}}\p{\lambda_{\bar 2}}{\cal G}_c-G(\bar 2 \bar 2^+;U)]\p{\lambda_{\bar 1}}{\cal G}_c
\right . \nonumber\\ && \left .
\mp 
[\p{\eta_{\bar 2^+}}\p{\lambda_{\bar 1}}{\cal G}_c-G(\bar 1 \bar 2^+;U)]\p{\lambda_{\bar 2}}{\cal G}_c
\right . \nonumber\\ && \left .
\p{\eta_{\bar 2^+}}{\cal G}_c[\p{\lambda_{\bar 2}}\p{\lambda_{\bar 1}}{\cal G}_c+\p{\lambda_{\bar 2}}{\cal G}_c\p{\lambda_{\bar 1}}{\cal G}_c]
\right \}.
\ee

Using (\ref{4}) we see that from the definition the correlated Green's functions
can be represented by functional derivatives of (\ref{invert}), 
\be
G_c(1 1';U)
&=&\!\!\left . \p{\eta_{1'}}\p{\lambda_1} {\cal G}_c\right
|_{\lambda=\eta=0}
\nonumber\\
&=&\!\!\left
    . \p{\eta_{1'}} {\cal I}(1) \right |_{\lambda=\eta=0}
\nonumber\\
G_c(12 1'2';U)
&=&\!\!\left . \p{\eta_{2'}}\p{\eta_{1'}}\p{\lambda_1}\p{\lambda_{2}}
  {\cal G}_c \right |_{\lambda=\eta=0}
\nonumber\\
&=&\!\!\mp \left . \p{\eta_{2'}}\p{\eta_{1'}}\p{\lambda_{2}} {\cal I}(1) \right |_{\lambda=\eta=0}
\nonumber\\
G_c(123 1'2'3';U)&=&
\!\!\left . \p{\eta_{3'}}\p{\eta_{2'}}\p{\eta_{1'}}\p{\lambda_1}\p{\lambda_{2}}\p{\lambda_{3}}
  {\cal G}_c \right |_{\lambda=\eta=0}
\nonumber\\
&=&\!\!\left
  . \p{\eta_{3'}}\p{\eta_{2'}}\p{\eta_{1'}}\p{\lambda_{2}}\p{\lambda_{3}}
  {\cal I}(1) \right |_{\lambda=\eta=0}
\label{int}
\ee
and so on. 
We drop the notation of the explicit dependence on the external potential U in the following. The first equation of (\ref{int}) yields for the one-particle correlated Green's function 
\ba
&G_c(11')=G_{\rm HF}(11')
\!\mp\! i \int d\bar 1 \bar 2 G_{\rm HF}(1\bar 1) V(\bar 1,\bar 2) G_c(\bar 1\bar 2 1'\bar 2^+)
\end{align}
and is nothing else but the Martin-Schwinger hierarchy (\ref{MS}) written for the correlated parts and in integral form.

\begin{figure}[h]
\centerline{ \psfig{file=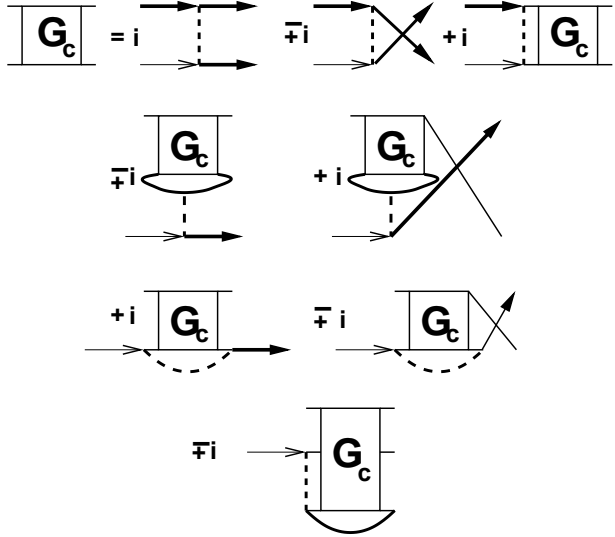,width=8cm}}
\caption{\label{as_parquet} The equation for the correlated two-particle
  Green's function. The thin open arrows are Hartree-Fock Green's functions, the thick
  arrows are full Green's functions. The first line represents the ladder approximation and the second and third line exchange channels.}
\end{figure}

The integral equation for the 
two-particle Green's function is more involved and reads from the second equation of (\ref{int})
\ba
&G_c(121'2')=i \int d\bar 1 d\bar 2 G_{\rm HF}(1\bar1)  V(\bar 1 \bar2)
\left . \biggl \{  \mp G_c(\bar 2\bar 1 2\bar 2^+1'2')
\right . \nonumber\\ & \left .
\!+\!G(2 \bar 2^+) \left [G_c(\bar 1 \bar 21'2')\!+\!G(\bar2 2') G(\bar 1 1')\!\mp\! G(\bar 2 1')G(\bar 1 2') \right]
\right . \nonumber\\ & \left .
+G_c(\bar 1 2\bar 2^+2')G(\bar 21')
+G_c(\bar 1 1'2\bar 2^+)G(\bar 22')
\right . \nonumber\\ & \left .
\mp G_c(\bar 2 2\bar 2^+2')G(\bar 11')
\mp G_c(\bar 2 2 1' \bar 2^+)G(\bar 12')
\right . \biggr \}.
\label{two}
\end{align}
This equation is graphical presented in figure \ref{as_parquet} and is exact
so far. Since the following algebra is somewhat involved it is easier to perform it in terms of diagrammatic presentation. Therefore we design a complete selfconsistent one-particle Green's function with a thick full arrow, the interaction with a broken line and the two-, three-, and four-particle Green's functions with a square containing two, three and four legs. The Hartree-Fock Green's function is designed as thin arrow. Corresponding upper signs denote fermions and lower signs bosons. The numbering of in and outgoing channels are in the direction from bottom to top. Lines without an arrow indicates ends of two- and three particle Green's functions which are not connected with a one-particle Green's function.

From the derivation above it is not hard to see that the property
\be
G_c(121'2';U)=\mp G_c(211'2';U)=\mp G_c(122'1';U)
\label{ex}
\ee
is ensured. In other words the criterion (A) and (B) of Kadanoff and Baym is fulfilled and we have a conserving approximation, i.e. the energy, momentum and density is conserved \cite{BK61}.
Please note that the corresponding diagrams in the second and third line in
figure \ref{as_parquet} cancel each other for fermions and contact
interaction, i.e. they become small for short-range interactions.

\section{Two-particle correlations}
\subsection{Binary collision approximation}
As a first very drastic approximation we want to consider the binary
correlation approximation which consists in neglecting
the correlated three-particle Green's function in (\ref{two}), i.e. the last
diagram  in figure (\ref{as_parquet}). Further we neglect also the second and third
line since they vanish for fermions and short-range interaction. This results
into the known ladder approximation or Bethe-Salpeter equation. Amazingly we have now obtained that the intermediate propagators of the Bethe-Salpeter equation have to be considered asymmetrically. Indeed, the vertex
\be
G_c(121'2')&=&\int d \bar 1 d\bar 2 d\bar 1' d\bar 2' G_{\rm HF}(1\bar 1)G(2 \bar 2) \Gamma (\bar 1 \bar 2 \bar 1' \bar 2')\nonumber\\&& \times 
G(\bar 1' 1')G(\bar 2'2'))
\ee
can be represented in symmetrized form 
\be
\Gamma(121'2')=T(121'2')\mp T(122'1')
\label{gamma}
\ee
by the T-matrix
\ba
&T(121'2')=iV(12)\delta(1-1')\delta(2-2')\nonumber\\&+i\int d\bar 1 d\bar 2 V(12) G(2\bar 2)G_{\rm HF}(1\bar 1) T(\bar 1 \bar 2 1' 2').
\label{tmatrix}
\end{align}

\begin{figure}[h]
\centerline{\psfig{file=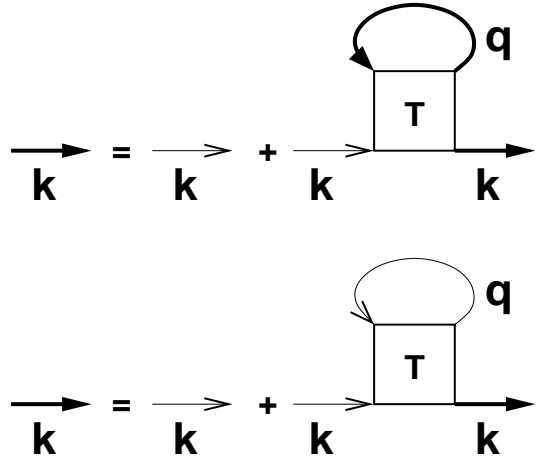,width=7cm}}
\caption{\label{tmat_dyson} The Dyson equation in ladder approximation for the symmetric selfconsistency (above) and the Kadanoff-Martin asymmetric one derived here (below).}
\end{figure}

The asymmetry
appears such that a full Green function $G_c$ is combined with a Hartree-Fock
Green's function. This asymmetry is maintained if we go systematically to
higher-order approximations neglecting 4-particle correlated Green's functions as
we will see in the next chapter. It is a result of the hierarchical structure
of the equations of motion. If one compares with the linearized parquet
approximations \cite{J01,J10}, the Hartree-Fock propagator $G_{\rm HF}$ in figure
(\ref{as_parquet}) is replaced by the full one $G_c$ which includes
nonphysical processes. We will see in the next chapter that the result derived here by the
hierarchy avoids such non-physically multiple scatterings with the same channel.
Therefore we consider the asymmetric form in figure \ref{as_parquet} as
superior to the parquet approximation.

In fact this asymmetry has been recognized as being necessary to obtain the gap equation for pairing \cite{Prange60}. It was first observed by Kadanoff and Martin \cite{KM61} and used later on \cite{MBL99,HCCL07,L08,SLMM09} as an ad-hoc approximation seemingly violating the symmetry of equations and consequently violating conservation laws. This has remained puzzling since a worse approximation leads obviously to better results. Recently it turned out that the repeated collisions \cite{L08} with the same particle are responsible for this artifact. 

\subsection{Derivation of Gap equation \label{gapc}}
Let us illustrate this by comparing the Dyson equation with symmetric and with asymmetric selfconsistency in figure \ref{tmat_dyson}. 
The T-matrix has poles at bound states and at the pairing and becomes
separable near these poles, $T=\mp\triangle\triangle$ where the upper sign is
for fermions and the lower for bosons. The two different Dyson equations for
the propagator with respect to the symmetry of selfconsistency is illustrated in the next figure \ref{tmat_nogap}. 
\begin{figure}[h]
\centerline{\psfig{file=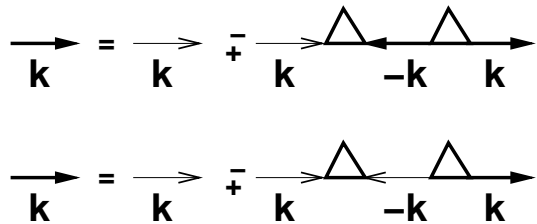,width=7cm}}
\caption{\label{tmat_nogap} The propagator equations resulting from figure \ref{tmat_dyson}, selfconsistently (above), Kadanoff-Martin asymmetric (below). The upper/lower sign stands for Fermi/Bose systems.}
\end{figure}
The first symmetrical selfconsistent Dyson equation leads to the propagator
\be
G(\omega,{\bf k})={1\over \omega-\epsilon_{\bf k}\pm \Delta^2G(-\omega,-{\bf k})} 
\ee
which shows no pole and no gap equation. Considering the second asymmetrical Kadanoff-Martin approximation of figure \ref{tmat_nogap} we obtain
\be
G(\omega,{\bf k})={1\over \omega-\epsilon_{\bf k}\mp{\Delta^2\over
    \omega+\epsilon_{-\bf k}}} 
\label{pole}
\ee
which possesses the typical two-pole structure of the BCS gap equation in the case of fermions. Therefore the Kadanoff and Martin approximation is superior to the symmetric selfconsistent ladder approximation and appears as a consequence of the hierarchical dependencies of correlations as derived above.
 
\begin{figure}[h]
\centerline{ \psfig{file=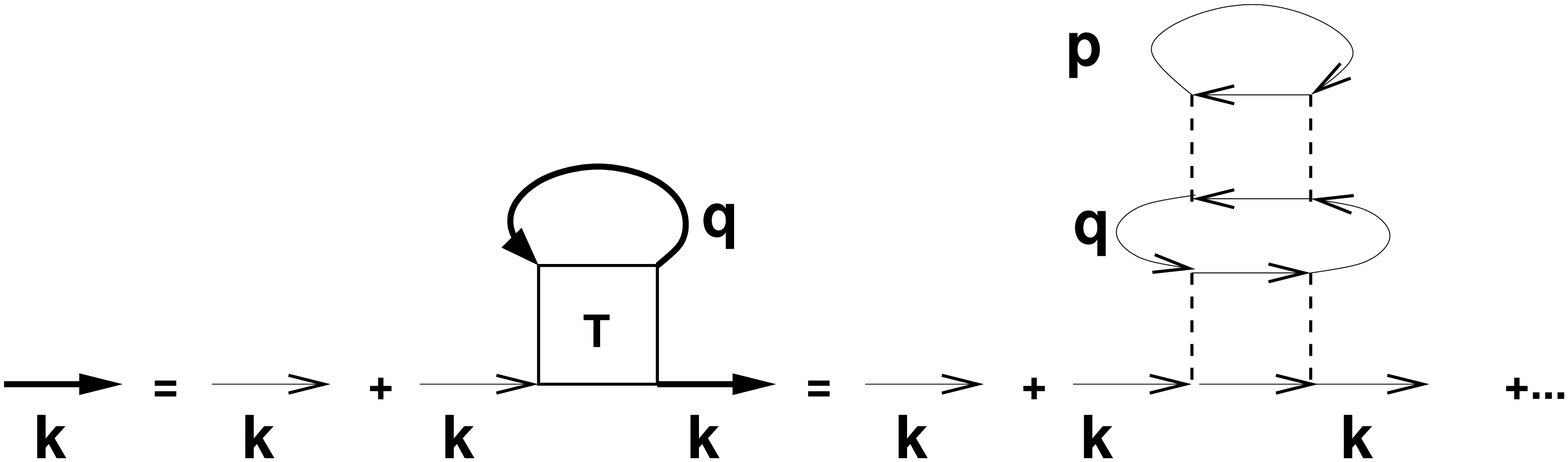,width=8.5cm}
}\caption{\label{tmat_iter} The iteration of the selfconsistent Dyson equation of figure \ref{tmat_dyson} leading to repeated collision with momentum ${\bf p, q}$.}
\end{figure}

What has this now to do with avoiding repeated collisions? Repeated collisions
of two particles in the same state are unphysical since the particles move
apart from each other after the collision. Therefore we have to ensure that
due to selfconsistency such collisions with the same state do not appear. In
fact as can be seen in figure \ref{tmat_iter}, the selfconsisent Dyson equation
does not ensure that the momentum of the repeated collisions, ${\bf p}$, is 
unequal to the momentum of the incoming
particle ${\bf k}$.  If this would be the case the particle will scatter with
a particle in the same state again. These repeated collisions have to be
removed from the T-matrix and the correct gap equation appears and the condensate can be described without asymmetrical ad-hoc assumptions about selfconsistency. 
The advantage of eliminating only the contributions of single channels as proposed in Refs.~\cite{L08} and \cite{SLMM09} is that the formation of pairs and their condensation can be described within the same approximation. 
Explicitly, we split the selfenergy into different channels, $\Sigma=\sum_j
\Sigma_j$ where we assume the condensation or pairing to appear in the channel
$i$ called the singular channel. Now we define a subtracted propagator 
\be
G_\s=G-G_\s\Sigma_iG
\label{sub}
\ee
where the T-matrix of the channel carrying the pole, $T_i$, is closed by the subtracted propagator $\Sigma_i=T_i\bar G_\s$. Writing the Dyson equation explicitly leads to the full propagator in momentum-energy representation $p=(\omega,{\bf p})$ \cite{M10}
\be
G(p)={\omega+\epsilon+\bar \Sigma_{11} \over (\omega+\epsilon+\bar \Sigma_{11})(\omega-\epsilon-\Sigma_{11})\mp\Sigma_{12}^2}
\label{polebose}
\ee
where $\Sigma_{11}(p)\equiv \Sigma(p)-\Sigma_i(p)$ and $\Sigma_{12}(p)\equiv
\Delta(p)$ and the upper/lower sign stands for fermions/bosons. The expression (\ref{polebose}) is nothing else but the Beliaev form for bosons \cite{B58}
or the Nambu-Gorkov form for fermions \cite{Gor58} and can be written also in matrix notation with off-diagonal elements \cite{M10}.

Though we have shown above that the Kadanoff-Martin approximation leads
already to a two-pole Green's function (\ref{pole}) necessary for the gap
equation, the alert reader has noticed that the subtraction of unphysical
multiple scattering events has
assumed a different intermediate propagator than the Kadanoff-Martin
approximation though leading to the same gap propagator (\ref{polebose}). The Kadanoff-Martin approximation
requires to use the Hartree-Fock propagator as derived above by approximating
the hierarchy of correlations at the binary level. For the subtraction of repeated
collisions we have used instead the  propagator written with the help of (\ref{sub})
\be
G_\s=G_{\rm HF}+G_{\rm HF}(\Sigma-\Sigma_{\rm HF}-\Sigma_i)G_\s
\label{gs}
\ee 
which shows that the subtracted propagator is beyond the Hartree-Fock one.
How can we derive such a form from the hierarchy of correlations?

\section{Three-particle approximations}
\subsection{Generic structure}
For this purpose we go one step further and approximate the equation for
the three-particle correlation represented by the last line of (\ref{int}). It
shows the coupling to the four-particle Green's function and is quite
lengthy. For the sake of legibility we
abbreviate the interchanges of indices of the corresponding foregoing
expressions by denoting them in the following formula 
within the same kind of brackets,
\ba
&G_c(1231'2'3')=\mp i \!\!\int\! d\bar 1 d\bar 2 G_{\rm HF}(1\bar1)  V(\bar 1 \bar2)
\!\biggl \{ G_c(23\bar 2\bar 1 \bar 2^+ 1'2'3')
\nonumber\\&
\mp G_c(2\bar 2\bar 1 1'2'3') G_c(3 \bar 2^+)
  \mp (2\leftrightarrow 3)
\nonumber\\&
+\left. \bigl \{
G_c(23\bar 2\bar 2^+2'3')
  G_c(\bar 1 1')\mp (\bar 1\leftrightarrow \bar 2)
\right .\nonumber\\& \left .
\mp G_c(2\bar 2\bar 2^+3')G_c(3\bar 1 1'2')
  \mp (2\leftrightarrow 3)
\right .\nonumber\\& \left .
+ [ \mp G_c(3 \bar 2^+) G_c(2\bar 2 2'3') G_c(\bar 1 1') +(\bar 1\leftrightarrow \bar 2)]\mp [(2\leftrightarrow 3)]
\right .\nonumber\\& \left .
+G_c(2 3\bar 2^+1') [G_c(\bar 2\bar1 2'3')+G_c(\bar 2 2')G_c(\bar 1 3')
\right .\nonumber\\& \left .
\mp G_c(\bar 2 3')G_c(\bar 1 2')] 
\right . \bigr \}
\nonumber\\&
\mp \{1'\leftrightarrow 2'\} 
+\{1'\leftrightarrow
  2'\leftrightarrow 3'\}
\biggr \}.
\label{g3}
\end{align}
Now we consider selected approximations by neglecting the four-particle
correlation and selecting special sets of diagrams. First let us show how the known channel approximations appear which are the screened ladder, the maximally crossed diagrams and the ladder diagrams. Then we will consider the pair-pair correlations and their influence on these three channels.
\begin{figure}[h]
\centerline{ \psfig{file=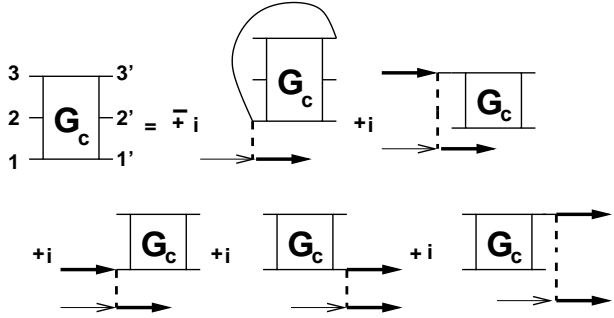,width=8cm}
}\caption{\label{vss} A special set of diagrams for the three-particle
  correlated Green's function (\ref{g3}).}
\end{figure}

\begin{figure}[h]
\centerline {\psfig{file=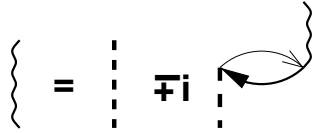,width=4cm}}
\caption{\label{screen} Definition of the screened potential $V_s$.}
\end{figure}

\subsection{Screened ladder approximation} 

\begin{figure}[h]
\centerline{ \psfig{file=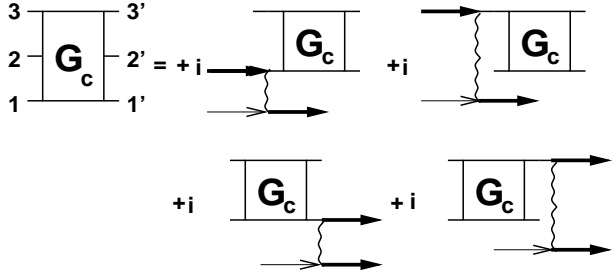,width=8cm}
}\caption{\label{gvs} The set of diagrams of figure \ref{vss} when introducing
  the screened potential of figure \ref{screen}.}
\end{figure}

\begin{figure}[h]
\centerline{ \psfig{file=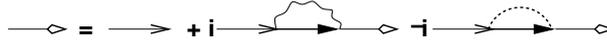,width=8cm}
}\caption{\label{screen_g} The screened propagator (\ref{mod}).}
\end{figure}

We choose as partial summation from (\ref{g3}) the first set of diagrams indicated in figure \ref{vss}.
The diagrams obtained by interchanging $1' \leftrightarrow 2'$ and $1'\leftrightarrow 2'
\leftrightarrow 3'$ in figure \ref{vss} are contained in (\ref{g3}) as well. Now the iteration of the equation for the three-particle correlated
Green's function in figure \ref{vss} leads to a repeated sum in the
interaction lines which can be summarized by introducing the screened
potential of figure \ref{screen}.
This procedure results into the expression for the three-particle correlated
Green's function as illustrated in figure \ref{gvs}.
\begin{figure}[h]
\centerline{ \psfig{file=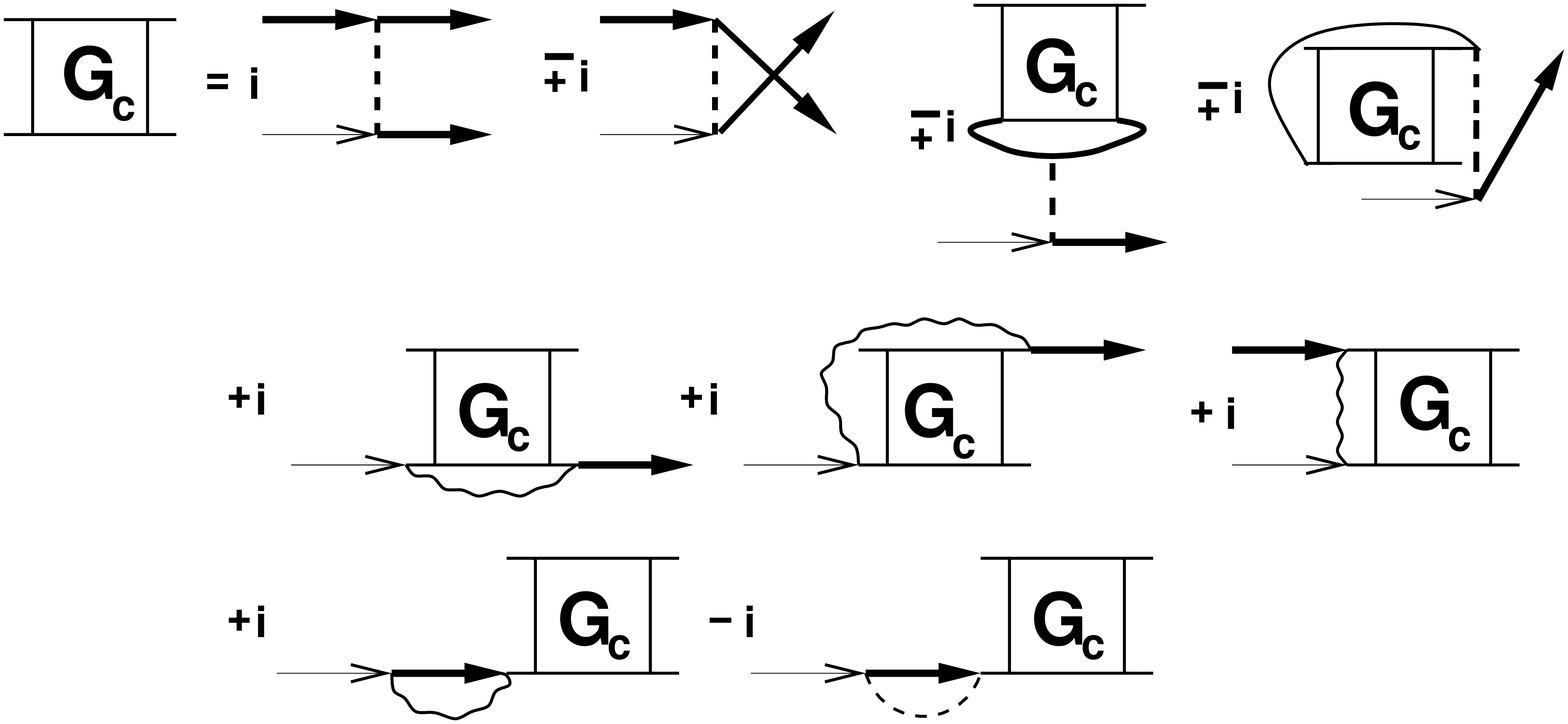,width=8cm}
}\caption{\label{screen_parquet} The diagrams of figure \ref{as_parquet} when introducing
  the three-particle Green's function of figure \ref{gvs}.}
\end{figure}
Introducing this expression into the last diagram of figure \ref{as_parquet}
one obtains the diagrams of figure \ref{screen_parquet}. We see that the last
line renormalizes the single-particle propagator if brought to the left-hand side.
In such a way the equation of the two-particle Green's function in figure \ref{screen_parquet}
can be very much simplified to which end we introduce the modified propagator
\be
G_s&=&G_{\rm HF}+G_{\rm HF}(\Sigma_s-\Sigma_{\rm F})G_s
\nonumber\\
&=&G_{\rm H}+G_{\rm H}(\Sigma_s)G_s
\nonumber\\
&=&G_{0}+G_{0}(\Sigma_s+\Sigma_{\rm H})G_s
\label{mod}
\ee 
where $\Sigma_s(12)=i V_s(12) G(12)$ is the screened self energy and
$\Sigma_{\rm F}(12)=i V(12) G(12)$ the Fock self energy as illustrated in figure \ref{screen_g}.
We will call in the following this modified propagator the channel-dressed selfconsistent propagator to distinguish him from the complete-dressed selfconsistent propagator. The channel-dressed propagator is determined by the corresponding selfenergy understood as the lowest selfconsistent diagram in the corresponding channel while the complete-dressed propagator includes all higher-order crossed terms. The introduction of this channel-dressed (screened) propagator results into the final expression of figure \ref{cs}.
\begin{figure}[h]
\centerline{ \psfig{file=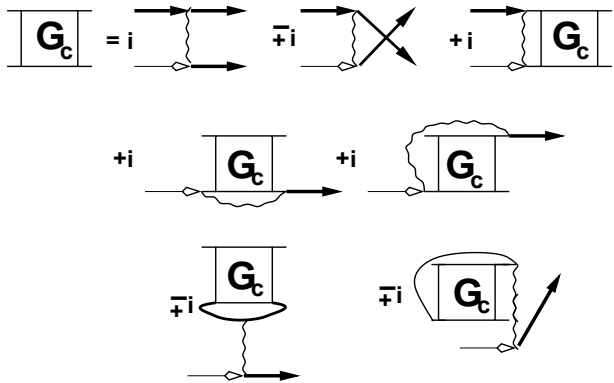,width=8cm}
}\caption{\label{cs} The diagrams of figure \ref{screen_parquet} when introducing
  the channel-dressed propagator (\ref{mod}) indicated as thin open arrows.}
\end{figure}

Compared to the Kadanoff-Martin approximation which was 
represented by diagrams of figure \ref{as_parquet} neglecting the three-particle
Green's function, we now obtain in figure \ref{cs} the same set of diagrams  except that the
Hartree-Fock propagator has to be replaced by the modified one (\ref{mod}) and
the bare interaction has to be replaced by the screened interaction of figure
\ref{screen} in the ladders.  Please note that the screened interaction
appears with asymmetric propagators.

Again we see that for fermions and short range
screened interaction the second and third line of diagrams in figure \ref{cs} are canceling
mutually and only the screened ladder diagram remains. Together with the
screened potential of figure \ref{screen} this is the screened ladder approximation used before \cite{zkkkr78} except that now one of the internal propagators has to be replaced by the screened one (\ref{mod}). This establishes the asymmetric form derived here as a new result.

Please note that (\ref{mod}) looks already like the structure of subtracting  unphysical
repeated collisions (\ref{gs}), however, only the Fock term appears to be
subtracted from the screened approximation selfenergy in the propagator
(\ref{mod}) since the scheme accounts for it already in the Bethe- Salpeter equation.

\subsection{Maximally crossed diagrams}

\begin{figure}[h]
\centerline{ \psfig{file=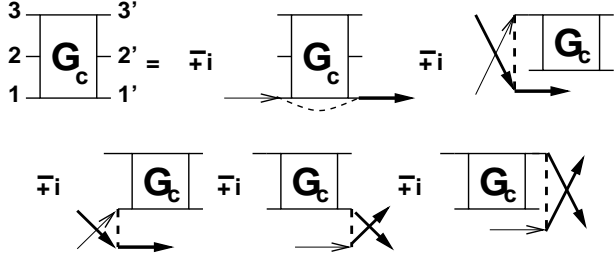,width=8cm}
}\caption{\label{vm} The maximally crossed diagrams contained in (\ref{g3}).}
\end{figure}

\begin{figure}[h]
\centerline{ \psfig{file=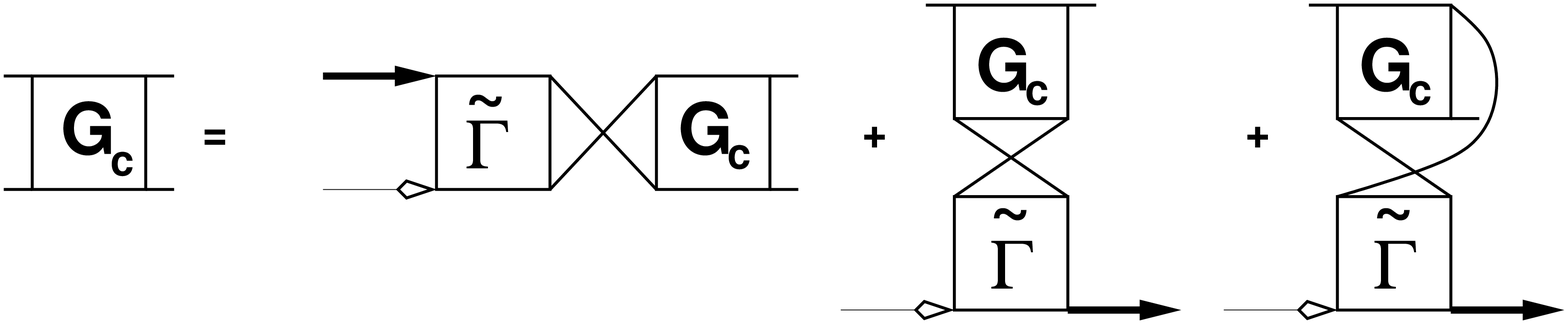,width=8cm}
}\caption{\label{cm} The diagrams introducing figure \ref{vm}  into figure \ref{as_parquet}. The thin open arrow mark now the channel-dressed propagator (\ref{modtt}).}
\end{figure}

\begin{figure}[h]
\centerline{ \psfig{file=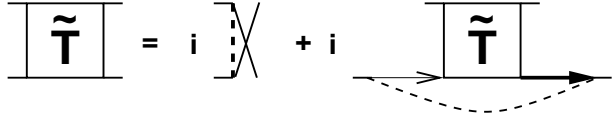,width=8cm}
}\caption{\label{tm} The maximally crossed ladder summation.}
\end{figure}

A next set of diagrams included in (\ref{g3}) is summarized by the maximally crossed ladders
presented in figure \ref{vm} where the diagrams interchanging $1' \leftrightarrow 2'$ and $1' \leftrightarrow 2'\leftrightarrow 3'$ are contained in (\ref{g3}) as well.
We proceed with the same steps as in the last chapter by introducing the
expression of figure \ref{vm} into the last diagram of figure \ref{as_parquet} but defining now the channel-dressed propagator
\be
G_{\tilde T}&=&G_0+G_0\Sigma_{\tilde T} G_{\tilde T}\nonumber\\
&=&G_{\rm HF}+G_{\rm HF}(\Sigma_{\tilde T}-\Sigma_{\rm HF}) G_T
\label{modtt}
\ee
to obtain the diagrams in figure \ref{cm}.
Here the selfenergy reads
\be
\Sigma_{\tilde T}(11')=\mp \int d \bar 1\bar 2 \,\tilde \Gamma(\bar 1 1\bar 2 1') G(\bar2 \bar1^+)
\ee
where the symmetrized vertex 
\be
\tilde \Gamma(1234)=\tilde T(1234)\mp \tilde T(1243)
\ee 
is expressed via the maximally crossed ladders of figure \ref{tm}.

\subsection{Ladder diagrams}

\begin{figure}[h]
\centerline{ \psfig{file=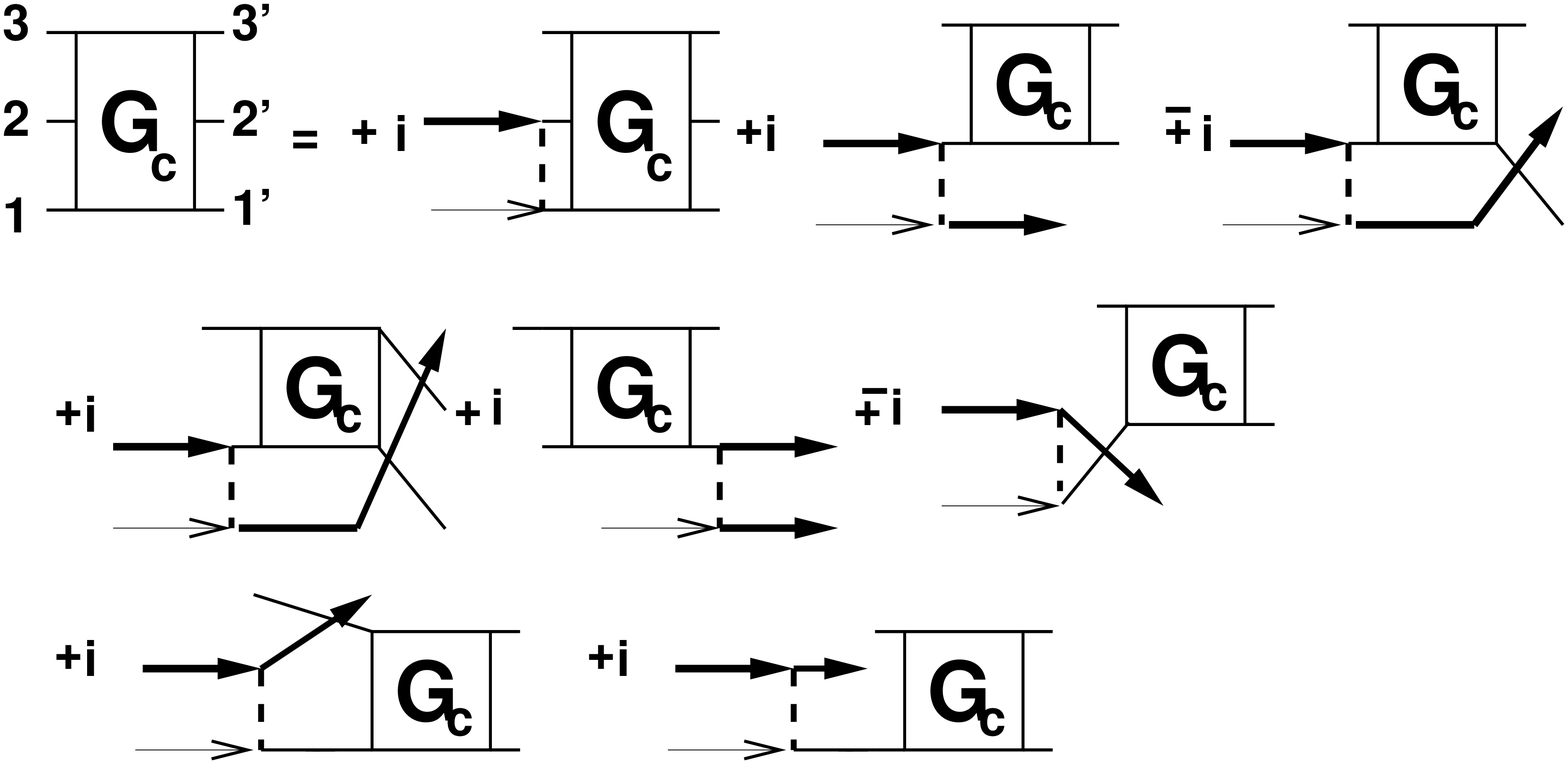,width=8cm}
}\caption{\label{vt} The ladder diagrams contained in (\ref{g3}).}
\end{figure}

As a last set we select the ladder diagrams included in (\ref{g3}) which are collected
in figure \ref{vt}. Also the diagrams interchanging $2 \leftrightarrow 3$  are contained in (\ref{g3}).

\begin{figure}[h]
\centerline{ \psfig{file=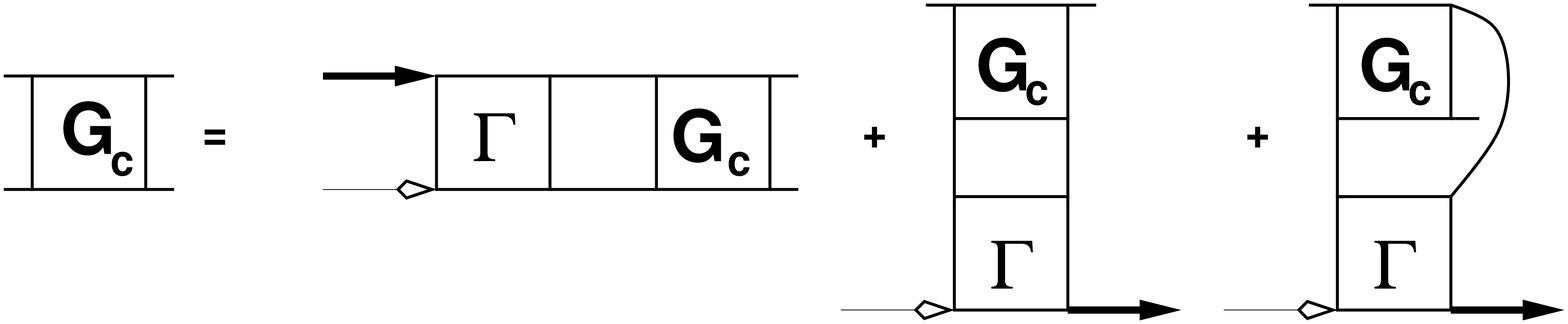,width=8cm}
}\caption{\label{ct} The diagrams introducing figure \ref{vt}  into figure \ref{as_parquet}. The thin open arrow are the channel-dressed propagator (\ref{modt}).}
\end{figure}

\begin{figure}[h]
\centerline{\psfig{file=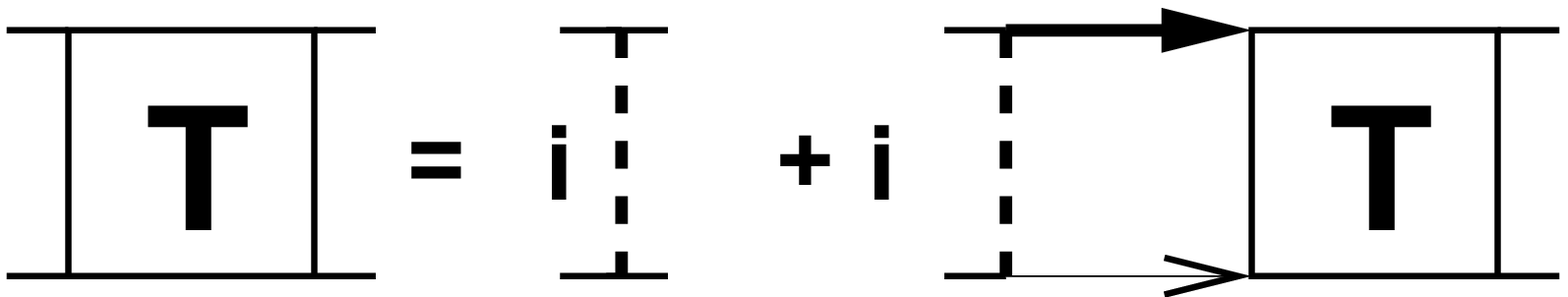,width=6cm}}
\caption{\label{t} The ladder summation.}
\end{figure}

Introducing the expression of figure \ref{vt} into the last diagram of figure
\ref{as_parquet} with the channel-dressed propagator
\be
G_{T}&=&G_0+G_0\Sigma_{T} G_T\nonumber\\
&=&G_{\rm HF}+G_{\rm HF}(\Sigma_{T}-\Sigma_{\rm HF}) G_T
\label{modt}
\ee
we obtain the diagrams in figure \ref{ct}.
Here the selfenergy reads
\be
\Sigma_{T}(11')=\int d \bar 1\bar 2 \,\Gamma(\bar 1 1\bar 2 1') G(\bar2 \bar1^+)
\ee
and the symmetrized vertex $\Gamma$ was given by (\ref{gamma})
in terms of the ladder T-matrix (\ref{tmatrix}) presented in figure \ref{t}.

\subsection{Pair-pair correlation}

\begin{figure}[h]
\centerline{ \psfig{file=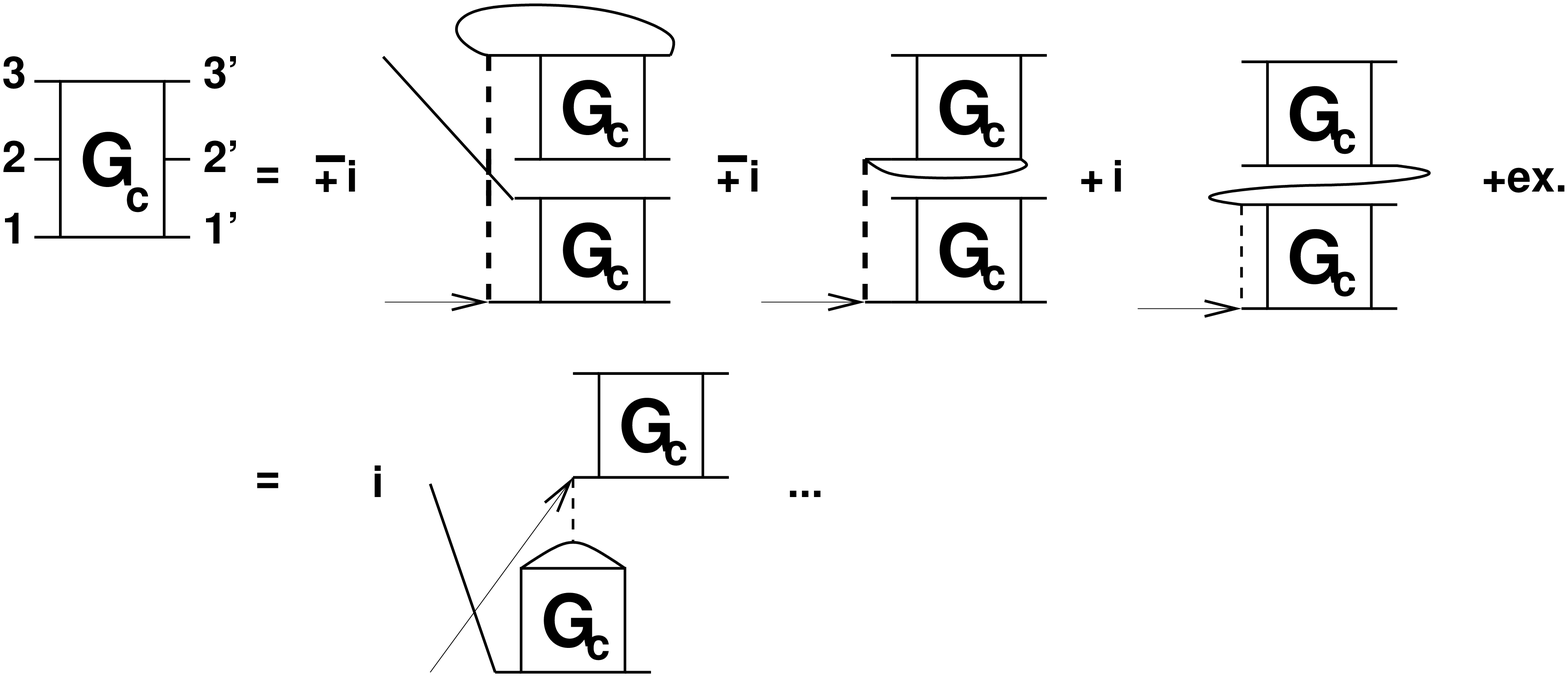,width=8cm}
}\caption{\label{vp} The pair-pair diagrams in (\ref{g3}).}
\end{figure}

Now that we have seen how the standard channels appear from the cummulant
expansion of correlations in terms of Green's functions with special emphasis on
the asymmetric propagators we proceed and investigate the pair-pair 
correlation. In fact in (\ref{g3}) there are diagrams included 
which describe the
interaction of two two-particle Green's functions outlined in figure \ref{vp}.

From these diagrams we search only the ones which yield a renormalization of
the left-lower Hartree-Fock propagator after iteration in one of the above three
channels, i.e. the first responsible diagram of figures \ref{vss}, \ref{vm} or \ref{vt}
respectively. In such a way we can define a channel-dressed propagator as done repeatedly
above. It turns out that only the one diagram written in the second line of
figure \ref{vp} fulfills this task. All other diagrams give repeated
iterations partially included in the above summations and partially leading to
new cross diagrams. These diagrams we will not consider here. They have been partially
considered in Born approximation named as cluster Hartree-Fock diagrams
\cite{R94} and have been applied to exciton problems \cite{RSSZ80} and to the
first-order superfluid phase transition \cite{R95}. These diagrams describe the interaction between the cluster and the
single particle, while we concentrate here on a genuine cluster-cluster diagram.

\begin{figure}[h]
\centerline{\psfig{file=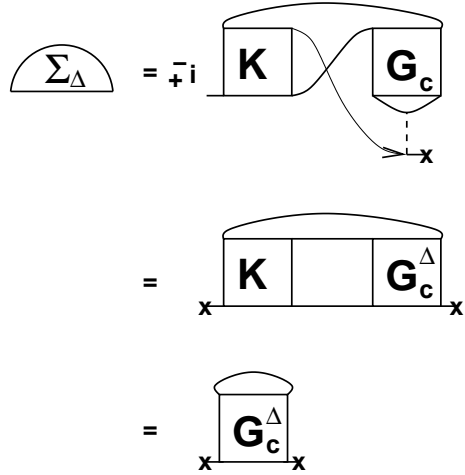,width=6cm}}
\caption{\label{sp} The selfenergy of the renormalized propagator due to
  pair-pair correlations in figure \ref{vp} where the kernel $K$ signs $V_s-V$
  for the screened channel of figure
  \ref{cs},
  $i\tilde T$ for the maximally crossed channel in figure \ref{cm} and $i T+V$ for the ladder diagrams in
  figure \ref{ct}. The second line is valid only for
separable two-particle correlations, see figure \ref{hp}, when using the
screened diagrams of figure \ref{screen_parquet}. The third line appears
using the ladder diagrams of figure \ref{ct}. Crosses denotes
the inverse (amputated) propagator.}
\end{figure}

We introduce now this single 
renormalizing diagram of the last line in figure \ref{vp} 
into  the corresponding three-particle ones on the right
hand sides of figures \ref{vss}, \ref{vm} or \ref{vt}. This leads to iterations which
sum the interactions and result into the channel effective blocks, screened
potential, maximally crossed vertex or T-matrix vertex. These diagrams are then
again introduced into the last diagram of figure \ref{as_parquet} as done
repeatedly before. 
In this way we obtain an additional renormalization diagram for the channel-dressed propagators of
(\ref{mod}), (\ref{modtt}) and (\ref{modt}) which can be written 
\be
G_{\Delta}&=&G_p-G_p\Sigma_\Delta G_\Delta
\ee
with $p=s,\tilde T,T$ denoting the channels. The corresponding selfenergy $\Sigma_\Delta$ due to the pair-pair
interaction is shown in figure \ref{sp} where the kernel $K$ denotes the considered channels $V_s-V$, $i\tilde T$ and $i T+V$ correspondingly.

We want to consider specially the particle-particle channel represented by the T-matrix since in this channel the pairing appears. There we have
\be
G_\Delta=
&=&G_{\rm HF}+ G_{\rm HF}(\Sigma_T-\Sigma_\Delta-\Sigma_{\rm HF}) G_\Delta\nonumber\\
&=&G_{0}+ G_{0}(\Sigma_T-\Sigma_\Delta) G_\Delta.
\label{gp}
\ee

\begin{figure}
\centerline{\psfig{file=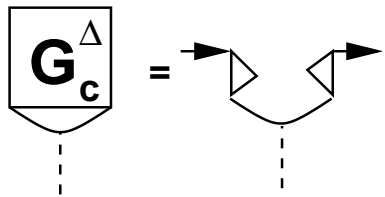,width=5cm}}
\caption{\label{hp} Separation in in- and outgoing channels in the pole of
  pairing or condensation.}
\end{figure}

For the singular channel, where the pairing appears, we will see now that the
desired 
specific contribution remains. For this purpose we use the fact that the
two-particle correlation separates near the pairing or condensation pole,
illustrated in figure \ref{hp} and as it was used above in chapter
\ref{gapc}. Taking this into account we see from the screened ladder diagrams
in figure \ref{screen_parquet} that besides the Hartree and Fock terms which
appear to be subtracted in figure \ref{sp}, only the third term in figure
\ref{screen_parquet} remains as a connected
diagram. Therefore we can replace the corresponding form in the second part of figure
\ref{sp} by a full two-particle propagator which results into the second line
of figure \ref{sp}. Using the iteration of the T-matrix channel of figure
\ref{ct} once more, we obtain the third line in figure \ref{sp}.
Due to this procedure we obtain the result   
\be
\Sigma_\Delta=\Sigma_\s
\ee 
and comparing (\ref{gs}) with(\ref{gp}) we concluded that
$G_\Delta=G_\s$.
This is exactly the subtracted propagator (\ref{gs}) proposed before which
corrects for the repeated collisions with the same state. The derived selfenergy
diagram in the first line of figure \ref{sp} is therefore one of the main results of this paper since it
presents the diagrammatic part which leads to the subtraction of unphysical repeated
collisions in any channel. For the particle-particle channel we have obtained the
correct subtracted propagator and see that a proper collection of
pair-pair correlations in the hierarchical expansion of correlations cares for the subtraction of such unphysical processes.

\section{Summary and outlook}

We have re-examined the Martin-Schwinger hierarchical structure of correlations. The expansion
scheme of Green's functions has been written with the focus on the cummulant
expansion form. We have investigated the binary collision approximation as well as
three-particle correlations and have derived the corresponding Bethe-Salpeter
equations. It appears that the two internal propagators of the resulting
Bethe-Salpeter equations are asymmetric with respect to
self-consistency. Dependent on the considered channel we obtain one complete-selfconsistent and one channel-dressed selfconsistent propagator. The
binary collision approximation leads to one Hartree-Fock and one
selfconsistent propagator in agreement with the Kadanoff and Martin form. This
leads to the gap equation of pairing while the standard Bethe-Salpeter
equation with two selfconsistent propagators does not. 

We have proceeded to
understand this asymmetry as a subtraction of repeated collision with the same
state which is unphysical. Such single-channel corrections contribute to diagrammatic sums with a 
weight of the inverse volume and appears therefore only if one has a singular
channel carrying a condensation with a macroscopic number of
occupation. It is shown that such a subtraction scheme appears if we consider the
correlations of two pairs of particles contained in the three-particle
correlated Green's function. We identify the selfenergy diagram which is
responsible for such proper subtraction scheme and show that the recently
proposed scheme following the idea of Soven has its justification in the
hierarchical dependence of correlations. 

The here derived selfenergy correction
responsible for this subtraction scheme  is valid in all three considered
channels, the screened ladder, the maximally crossed diagrams and the T-matrix
channel. Though we had searched only for the correction of the latter one in
order to obtain pairing and the gap equation we suggest that this selfenergy diagram might have an effect also
to the maximally crossed diagrams. In that channel one describes weak
localization phenomena. Since it was shown recently \cite{J10} that the
parquet diagrammatic summation leads not to the expected Anderson 
localization, we
suggest that the asymmetric parquet summation which appears due to the
here derived selfenergy correction might solve these problems with localization. This investigation
is deserved for further work.

\begin{acknowledgements}

The discussions with Pavel Lipavsk\'y, Bretislav \v Sop\'ik, Michael M\"annel
and Alvaro Ferraz are gratefully mentioned.
This work was supported by DFG-CNPq project 444BRA-113/57/0-1 and the DAAD-PPP
program (BMBF). The financial support by the Brazilian Ministry of Science 
and Technology is acknowledged.
\end{acknowledgements}

\appendix

\section{Variational technique, Ward identities and $\Phi$-derivability}

\begin{figure}[h]
\centerline{ \psfig{file=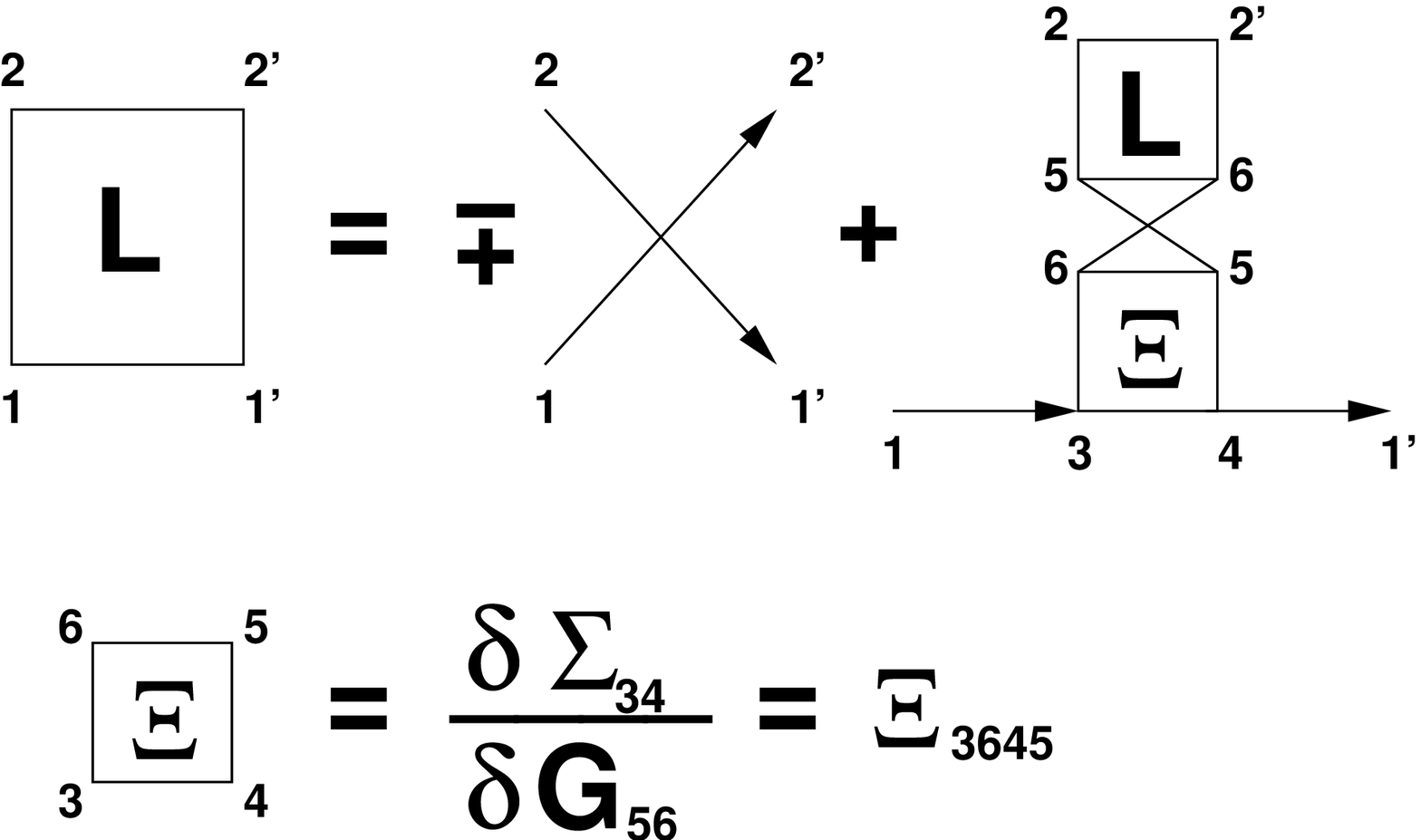,width=8cm}
}\caption{\label{v1}Variational representation of the two-particle correlation  function (\ref{3a})}\end{figure}

Due to the coupling of an external potential $U(11')=\delta_{11'}U(1)$ 
one expresses the two-particle causal Green function $G(121'2')=1/i^2 <T\Psi_1\Psi_2 \Psi_2^+\Psi_1^+>$ by a variation of the one-particle Green's function $G(12)=1/i<T\Psi_1\Psi_2^+>$ with respect to the external potential \cite{KB62,kker86} as
\be
G(121'2')=G(11')G(22')\mp {\delta G(11')\over \delta U(2'2)}
\label{1a}
\ee
where the upper sign denotes the Fermi and the lower the Bose functions.
Using the Dyson equation
\be
G^{-1}=G_0^{-1}-\Sigma-U
\label{2a}
\ee 
one can calculated the derivative in (\ref{1a}) and
with the help of the chain rule and $\delta G=-G \delta G^{-1} G$, 
the fluctuation function reads
\be
L(121'2')&=&G_c(121'2')+G(12')G(21')\nonumber\\
&=&G(121'2')-G(11')G(22')\nonumber\\
&=&\mp G(12')G(21')\mp G(13){\delta \Sigma(34)\over \delta U(2'2)} G(41')
\nonumber\\
&=&\!\!\mp \!G(\!12')G(\!21')\!+\!G(\!13) {\delta \Sigma(\!34)\over \delta G(\!56)} L(\!5262')G(\!41').
\nonumber\\&&
\label{3a}
\ee
Double occurring indices are understood as integrated over.
Equation (\ref{3a}) is expressed graphically in figure \ref{v1}.

\begin{figure}
\centerline{ \psfig{file=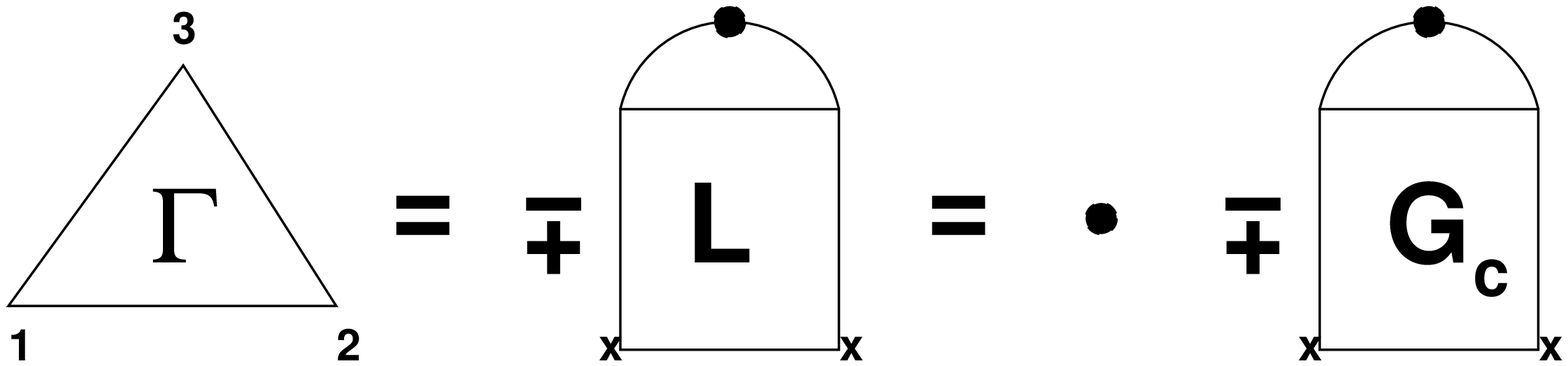,width=8cm}
}\caption{\label{v101} Definition of the vertex function (\ref{13})}\end{figure}

\begin{figure}
\centerline{ \psfig{file=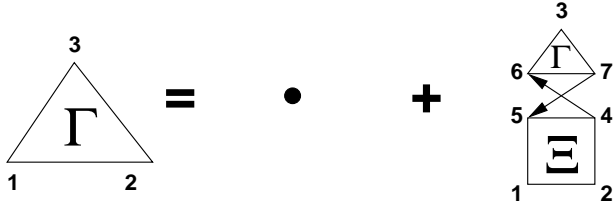,width=8cm}
}\caption{\label{v11}Equation for the vertex function.}\end{figure}

Defining a three-point vertex function
\be
\Gamma(123)&=&\delta_{12}\delta_{13}+{\delta {\Sigma}(12)\over \delta {U}(33)}\nonumber\\
&=&\delta_{12}\delta_{13}+{\delta {\Sigma}(12)\over \delta G(45)} G(46) \Gamma(673) G(75)
\label{13}
\ee
illustrated in figure \ref{v101}
the two-particle correlation function of figure \ref{v1} can be represented by figure \ref{v11}

The first line of equation (\ref{13}) establishes the Ward identity \cite{Ward51,T57,VKS08} 
which is therefore a consequence of the variational expression (\ref{1a}) and
the form of Dyson equation (\ref{2a}).  This Ward identity is also a
consequence of the fact that the selfenergy can be represented as variation of
a $\Phi$-functional with respect to the propagator \cite{MS59}. Recently it
has been proposed \cite{LDS06} to expand the variational scheme in terms of the kernel
represented in the last line
of figure \ref{v1}.
\begin{figure}[h]
\centerline{ \psfig{file=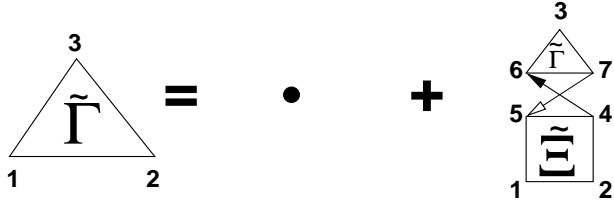,width=8cm}
}\caption{\label{v111}Equation for the vertex function in asymmetric
  form.}\end{figure}
\begin{figure}[h]
\centerline{ \psfig{file=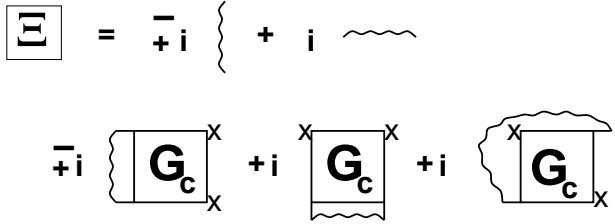,width=8cm}
}\caption{\label{kern}Kernel of the vertex function in figure \ref{v11} for screened ladder channel of figure \ref{cs}.}\end{figure}

How does this scheme fit now the asymmetric form presented in this paper? We
demonstrate this on the example of the screened ladder approximation of figure
\ref{cs}. The other channels can be treated analogously. Closing the upper
lines of the mirrored diagrams of figure \ref{cs} with the channel-corrected
propagator we obtain the same form of the three-point vertex in figure 
\ref{v11} but now translated into figure \ref{v111} with asymmetric
propagators. 
The kernel  takes the form represented by figure \ref{kern} which  
can be verified by expanding the diagrams with the help of $G_c$ of figure
\ref{cs}. In comparison with figure \ref{v11} the intermediate propagators are
asymmetric.

\begin{figure}[h]
\centerline{\psfig{file=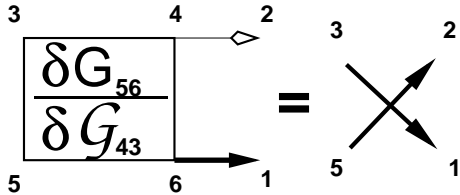,width=6cm}}
\caption{\label{reldef} Definition of the variational propagator ${\cal G}$ (\ref{rel2}).}\end{figure}

Using the Dyson equation for the channel-dressed propagator
with channel index $i$ we can
replace in one propagator of figure \ref{v11} 
\be
G=G_i+G_i\Sigma_iG
\ee
and then have to show that 
\be
\Xi(1234)+\Xi(1235)G_i(56)\Sigma_i(64)\equiv \tilde \Xi(1234)
\label{rel1a}
\ee
in order to proof $\tilde \Gamma=\Gamma$. To complete (\ref{rel1a}) 
we use $1+G \Sigma_i=GG_i^{-1}$ and write 
\be
\tilde \Xi(1234)&=& \Xi(1235)G_i(56)G_i^{-1}(64)\nonumber\\
&\equiv& 
{\delta \Sigma(12)\over \delta {\cal G}(43)}
={\delta \Sigma(12)\over \delta {G}(56)}{\delta G(56)\over \delta {\cal G}(43)}
\label{rel1}
\ee
where we have introduced the (variational) selfconsistent propagator ${\cal
  G}$ to be defined by
\be
{\delta G(56)\over \delta {\cal G}(43)}
G_i(42)G(61)\equiv G(31)G(52)
\label{rel2}
\ee
illustrated in figure \ref{reldef}.

With the help of this variational propagator we can
fulfill the relation (\ref{rel1}) such that we have
\be
\tilde \Xi(1234)= 
{\delta \Sigma(12)\over \delta {\cal G}(43)}.
\label{rel3}
\ee
This, however, has the required form of figure \ref{v1} such that
(\ref{3a}) is fulfilled and the Ward identities are completed. Please note
that neither $G$ nor $G_i$ serve as propagators to perform the variation but
the variational propagator ${\cal G}$ defined by (\ref{rel2}).

Therefore the screened ladder approximation (and the other 2 considered
channels as well) can be recast into a form of three-point vertex function
which obeys an integral equation with has the desired form of Ward
identity. The difference to the traditional treatment is that a full dressed
propagator is combined with the channel-dressed propagator. The price we pay
is that we do have two different selfconsistent propagators the
channel-dressed one and the complete dressed one.

Now we can write down the standard $\Phi$-functionals for T-matrix, screened
ladder and maximally crossed channels but with symmetrical ${\cal G}$
propagators. 

\begin{figure}[h]
\centerline{\psfig{file=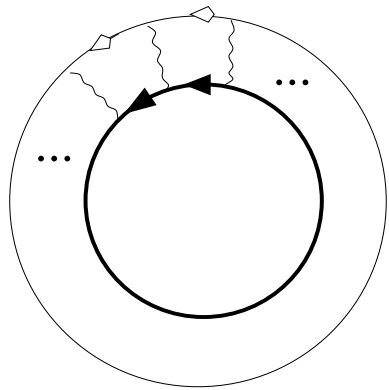,width=5cm}}
\caption{\label{phi} The $\Phi$-functional defining the asymmetric screened
  channel approximation.}\end{figure}

Alternatively we can use (\ref{rel2}) to replace them by the two
different selfconsistent propagators.
Consequently, the appearance of two different propagators does not violate the
$\Phi$-derivability and the conservation laws. This can be considered
alternatively by translating the known conserving approximations into forms of two different species in the system and interpret in the end one specie as the complete-dressed selfconsistent and the other specie as the channel-dressed selfconsistent propagator. 
For the screened ladder channel we declare e.g. a $\Phi$-functional as illustrated in figure \ref{phi}.

At the end it has to be remarked that the here proven Ward identity and $\Phi$-derivability is in a sense a triviality as
long as the scheme is kept exact. Any formal expansion is 'exact' and will
obbey therefore these exact identities. The difference between different
expansion schemes lays in the different kind of approximations 'naturally'
offered. Therefore at any stage of approximation actually used, one has
to ensure the Ward identity. The introduction of the variational propagator
(\ref{rel2}) in figure \ref{reldef} shows that any asymmetric approximation
proposed with the scheme in this paper can be translated into a symmetric
one. However the relation is very cumbersome in such a way that any asymmetric
form chosen translates into many higher order diagrams in the symmetric writing.

\bibliography{bose,kmsr,kmsr1,kmsr2,kmsr3,kmsr4,kmsr5,kmsr6,kmsr7,delay2,spin,refer,delay3,gdr,chaos,sem3,sem1,sem2,short,cauchy,genn,paradox}
\bibliographystyle{spphys}       

\end{document}